\documentclass[useAMS,usenatbib]{mn2e}
\usepackage[T1]{fontenc}

\usepackage{ae,aecompl}
\usepackage{times}
\usepackage{url}
\usepackage{graphicx}
\usepackage{graphics}
\usepackage[usenames]{color}
\DeclareGraphicsExtensions{.pdf,.png,.jpg,.mps,.eps,.ps}
\usepackage{amsmath}
\usepackage{natbib}

\def\unit #1{\,{\rm #1}}

\newcommand\kms{\rm \,\unit{km\,s^{-1}}}

\newcommand\cmsqi{\rm \,\unit{cm^{-2}}}

\newcommand\kev{\rm \,\unit{keV}}

\newcommand\funit{\rm \,erg\,cm^{-2}\,s^{-1}}

\newcommand\xiunit{\rm \,erg\,cm\,s^{-1}}

\newcommand\nh{\rm N_{H}}

\newcommand\nhwa{N^{\rm WA}_{\rm H}}

\newcommand\ks{\, \rm ks}

\newcommand\dc{\, \Delta\chi^2}

\newcommand\cd{\,\rm \chi^2/dof}

\newcommand\ev{\unit{\, eV}}
\newcommand{\caxx}{\mbox{Ca\,{\sc xx}}}

\newcommand\chandra{{\it Chandra}}

\newcommand\xmm{{\it XMM-Newton}}

\usepackage{ulem}
\usepackage{graphicx}

\title{On the origin of the {featureless} soft X-ray excess emission from the
  Seyfert 1 galaxy ESO~198--G24.}

\author [Laha et al.] {Sibasish laha$^{1}$\thanks { Email: laha@iucaa.ernet.in},
Gulab \ C.\ Dewangan$^{1}$, Ajit \ K.\ Kembhavi$^{1}$ \\ $^{1}$  Inter University Center for Astronomy and Astrophysics, Pune , India}

\date{\today}
\begin{document}
\pagerange{\pageref{firstpage}--\pageref{lastpage}} \pubyear{2013}




\maketitle
\label{firstpage}
\begin{abstract}

  We present medium and high resolution X-ray spectral study of a
  Seyfert 1 galaxy ESO~198--G24 using a long ($\sim 122 \ks$) \xmm{}
  observation performed in February 2006. The source has a prominent featureless soft X-ray excess below
  $2\kev$. This makes the source well suited to investigate the origin of the soft excess. Two physical models -- blurred reflection, and optically thick thermal Comptonization in a warm plasma, describe the soft-excess equally well resulting in similar fits in the $0.3-10\kev$ band. These models also yield similar fits to the broad-band UV (Optical Monitor) and X-ray data. \xmm{} observations performed in 2000, 2001 and 2006 on this source show flux variability. From 2001 to 2006, the UV flux increased by $\sim23\%$ while the $2-10\kev$ X-ray flux as well as the soft-excess flux decreased by $\sim 20\%$. {This observation can be described in the blurred reflection scenario by a truncated accretion disk whose inner-most radius had come closer to the blackhole. We find that the best-fit inner radius of the accretion disk decreases from $\rm R_{in}=4.93_{-1.10}^{+1.12}R_G$ to $\rm R_{in}<2.5R_G$ from 2001 to 2006.} This leads to an increase in the UV flux and compressing the corona, leading to reduction of the powerlaw flux and therefore the soft-excess. The blurred reflection model seems to better describe the soft-excess for this source.

\end{abstract}


\section{Introduction}

Many Seyfert 1 galaxies show a prominent soft X-ray excess (SE)
emission over and above a powerlaw component extending to high
energies. This SE was first detected by
  \cite{1981ApJ...251..501P} using {\it HEAO-1} data and also by
  \cite{1985MNRAS.217..105A} and \cite{1985ApJ...297..633S} using {\it EXOSAT}
  data. The nature and origin of the SE is still uncertain. In most
cases it is well described by blackbody or
multiple blackbody emission with photon temperature ranging from
$0.1-0.2 \kev$ over several decades in AGN mass
\citep{2006MNRAS.371L..16G}. If one assumes this feature to have a
thermal origin, then its temperature is quite high as compared to that
predicted by the standard accretion disk model of
\cite{1973A&A....24..337S}. The constancy of the temperature of the SE
points to the fact that its origin is likely related with the atomic
physics, in a way that the shape of the SE is the result of atomic
transitions. The SE can be described as blurred reflection
from a partially ionised accretion disk when a hard X-ray photon flux
is incident on it \citep{2005MNRAS.358..211R}. It can also be
physically explained by Comptonisation of disk photons by an optically
thick electron cloud at higher temperature
\citep{2012MNRAS.420.1848D}. The soft excess is usually modified by
the presence of strong and complex warm absorber features. This makes the
characterization of both the soft excess and the warm absorbers
difficult particularly if the spectral resolution is moderate or the
signal to noise ratio is low. Therefore, it is important to study Seyfert 1
galaxies without significant soft X-ray warm absorber or other features in order to
probe the nature of SE emission.  The Seyfert 1 galaxy
ESO~198--G24 is one such AGN with featureless soft X-ray excess.

ESO~198--G24 was observed in the soft X-ray band by {\it
  ROSAT} twice during $1991-1992$ \citep{1993ApJ...412...72T}. It was
noted that the spectrum had varied between the two observations. There
was a flattening of the soft X-ray spectrum, and the emission lines and
absorption edges were not required statistically in the second 1992
observation. It was possible to put a lower limit on the variability
time scale of six months. 

\cite{2003A&A...401..903G} studied ESO~198--G24 using the $\sim 40
\ks$ data from {\it ASCA} (1997), $\sim 9\ks$ data from {\it
  XMM-Newton} (2000), and $\sim150\ks$ from {\it BeppoSAX} (2001). The
study revealed that the Fe K$\alpha$ line profile and the line
intensity varied between the observations. In the {\it XMM-Newton}
data, \cite{2003A&A...401..903G} found an additional emission line
feature at $5.7 \kev$ with an equivalent width $70\pm 40 \ev$ and
suggested that it might be a part of the double horned profile of the Fe line.

\cite{2004A&A...413..913P} studied the source using an {\it
  XMM-Newton} observation performed in 2001. There was no clear
evidence of warm absorption in the soft X-ray band, however there were
weak relativistic emission lines of OVIII and CVI Ly$\alpha$ in the
RGS data. An FeK$\alpha$ line at $6.41 \kev$ with an equivalent width
of about $60-70 \ev$ was clearly detected.  As a part of a sample
study, \cite{2010A&A...521A..57T} have investigated the presence of high
ionisation outflows from ESO~198--G24 using data obtained from an
{\it XMM-Newton} observation in 2006. They did not find any narrow
absorption line in the Fe K band, however they found a narrow
absorption edge at $4.59 \kev$. It could possibly be associated with
\caxx~Ly~$\alpha$, blueshifted by a velocity of $\sim 0.1 $c.
\cite{2009A&A...507..159D} studied the source as a part of a sample
study investigating the variability of the Fe K complex. They used the
time averaged spectrum from $4-9\kev$, and found that the Fe K$\alpha$
line varied between the two {\it XMM-Newton} observations in 2001 and
2006, in that the former data showed a broader Fe K$\alpha$ line with
an upper limit on the line width $\sigma < 0.36 \kev$. They also
found an absorption line in the 2006 {\it XMM-Newton} data at $7.56
\kev$ with an equivalent width of $40 \ev$.

{ In this paper we make a detailed broadband spectral study of
ESO~198--G24 using all the available {\it XMM-Newton} datasets and
investigate the origin of the soft X-ray excess emission. This is the first detailed spectral analysis for this long $122$ ks dataset. Section 2
describes the observations and data reduction. Section 3 deals with
broad band spectral analysis (EPIC-pn and OM+EPIC-pn),
as well as RGS spectral analysis. We discuss our results in section 4
followed by the conclusions.}

\begin{table}
{\footnotesize
\centering
  \caption{The three \xmm{} observations of ESO~198--G24 used in this work.  \label{obs}}
  \begin{tabular}{l l l l llll} \hline\hline 

Observation & Observation & Observation & Total exposure& \\ 
   number   &    id          &date               &  time\\  \hline 

1 &0112910101  & 2000-12-01 & 13 ks &\\
2&0067190101   & 2001-01-24 & 34 ks  & \\
3 &0305370101  &2006-02-04  & 122 ks  & \\ 

\hline \hline
\end{tabular} \\ 
}
\end{table}

\section{Observation \& Data Reduction}

ESO 198--G24 was observed by {\it XMM-Newton} on three occasions in
December 2000, January 2001 and February 2006 for $13$, $43$ and
$122\ks$, respectively {(see Table \ref{obs} for details). We have analysed in detail the data obtained from the long $122\ks$ observation on 2006 (observation number 3, see Table \ref{obs}). In this observation, the
EPIC-pn camera was operated in the small window mode and the MOS
cameras were operated in the partial window mode.} Data were reduced using SAS version 11.0.0.

\begin{figure}
  \centering

  \includegraphics[width=7cm,angle=-90]{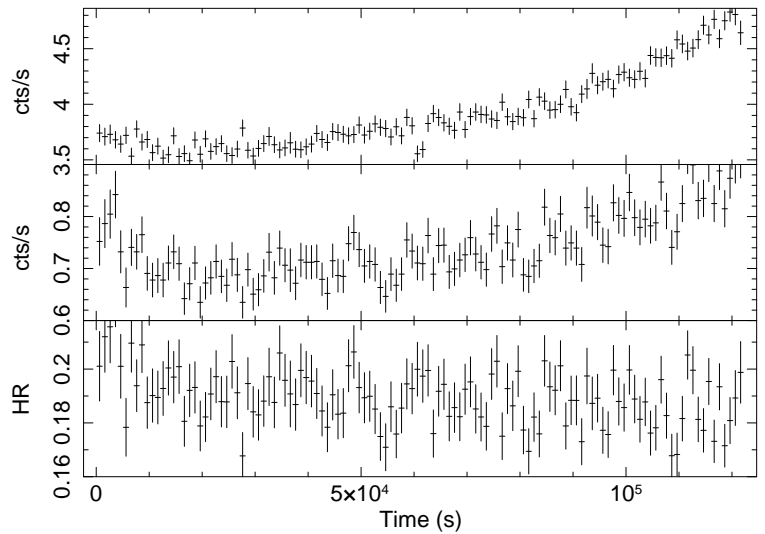}
  \caption{Background subtracted EPIC-pn lightcurves of ESO~198--G24 in
    the soft ($0.2-2\kev$) band (top panel), in the hard ($2-10\kev$)
    band (middle panel), and the hardness ratio (bottom panel) for observation 3 (id:0305370101).}.
  \label{hardness}
\end{figure}

We processed the EPIC-pn and MOS data with {\it epchain} and {\it
  emchain} respectively, and filtered them using the standard
filtering criterion.  Examination of the background rate above 10 keV
showed that the observation was partly affected by a flaring particle
background at the beginning of the observation upto an elapsed time of
$5 \ks$. We checked for the
photon pile-up using the SAS task {\it epatplot} and found that there
were no noticeable pile-up in either EPIC-pn or MOS data. We quote
results based on EPIC-pn data due to its higher signal-to-noise
compared to the MOS data. We have used single pixel events with
pattern=0 and FLAG=0. To extract the source spectrum and lightcurve,
we chose a circular region of $43{\rm~arcsec}$, centred on the
centroid of the source. For background spectrum and lightcurve, we
used nearby circular regions that are free of any sources.  Figure
\ref{hardness} shows the background-corrected light curves of the
source in the soft ($0.2-2 \kev$) and hard X-ray ($2-10 \kev$) bands
along with the hardness ratio. We used the full length of the
observation to extract the lightcurves. The net source light curve was
obtained after appropriate background subtraction using FTOOLS task
{\it lcmath}. We note that the hardness-ratio has not varied significantly within the observation.
 We created the ancillary response file (ARF) and the
redistribution matrix file (RMF) using the SAS tasks {\it arfgen} and
{\it rmfgen}, respectively. After filtering, the net exposure in the EPIC-pn data is
$85\ks$, and the net counts is $4 \times 10^5$.

We processed the RGS data using the SAS task {\it rgsproc}. We chose a
region, CCD9, that is most susceptible to proton events and generally
records the least source events due to its location close to the
optical axis and extracted the background light curve. We then
generated a good time interval file to filter the eventlist and
extracted the first order source and background spectrum.

We processed the optical monitor (OM) data using {\it omichain}. We
obtained the count rates in 5 active filters by specifying the RA and
Dec of the source in the source list file obtained by the {\it
  omichain} task. The count rate thus obtained was converted to flux
using standard tables. See section \ref{sub:OM} for details.
 For variability studies (section \ref{sub:variability}), we have also analysed the OM and EPIC-pn datasets from two earlier \xmm{} observations, for which we follow the same procedure as described above for data reduction. There was no pile up in the data for these observations. The hardness ratio of the source was found to be constant during these two observations.

\section{Spectral Analysis}

We grouped the EPIC-pn spectral data with a minimum of 20 counts per
energy bin and at most 5 energy bins per resolution element, using the
{\it specgroup} command in the SAS. We used ISIS version 1.6.1-36
\citep{2000ASPC..216..591H} for our spectral fitting. All errors
quoted on the fitted parameters reflect the $90\%$ confidence interval.

\begin{figure}
  \centering
  \includegraphics[width=6cm,angle=-90]{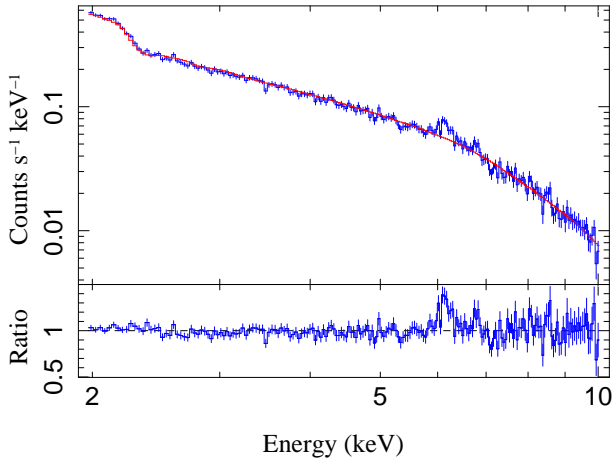}
  \caption{ The $2-10 \kev$ EPIC-pn data fitted by a simple absorbed
    powerlaw model to show the Fe K line complex in the energy range
    of $6-8 \kev$ (observer's frame). FeK$\alpha$ and K$\beta$ emission lines and weak
    FeK absorption lines are visible in the residuals.}
  \label{absorb}
\end{figure}

We begin with the spectral analysis of the $2-10 \kev$ data. A
simple powerlaw model with absorption due to neutral column in our
Galaxy (wabs) provided a $\rm \chi^2/dof= 292/198 \sim 1.50$, where
dof stands for degrees of freedom. The best fit equivalent neutral
Hydrogen column density, $\rm N_{H}\le 3 \times 10^{20} \cmsqi$, is
consistent with the Galactic column \citep[$\rm N_H^{Gal} = 2.93 \times
10^{20} \cmsqi$; ] []{2005A&A...440..775K}, and therefore we fixed
$\rm N_H$ to this value. There was no evidence for an additional
neutral absorber intrinsic to the source. The best fit powerlaw slope was
$\Gamma= 1.64_{-0.01}^{+0.04}$. Figure \ref{absorb} shows the $2-10\kev$ spectrum
fitted with an absorbed powerlaw model, and the
ratio of the observed data and the model. There are two emission lines
at $\sim 6.4 \kev$ and $\sim 7\kev$ in the rest frame ($z=0.0455$) and
also there are weak absorption features at $\sim 7.4 \kev$. The
Fe~K$\alpha$ feature has been detected previously in the same dataset
as a part of sample study by \cite{2010A&A...521A..57T} and
\cite{2009A&A...507..159D}.

We used a Gaussian profile to fit the prominent Fe~K$\alpha$ emission
line at $6.4 \kev$. {The best fit line energy is $6.41_{-0.05}^{+0.04}\kev$}.  Initially we fixed the {standard deviation of the Gaussian line, $\sigma$,} to a
value $0.01 \kev$ and found a weak red wing with $\cd= 229/194$. Then we allowed the $\sigma$ to
vary and the fit improved to $\cd=213/193$. We found that the Gaussian
line was slightly broader than the instrumental resolution with $\sigma = 0.12_{-0.04}^{+0.01}\kev$,
which corresponds to an FWHM speed of $13000_{-4250}^{+1000} \kms$. The
equivalent width of the line is $ 78\pm 20 \ev$. We then checked for a
possible relativistically broadened iron line component around $6
\kev$ with an ISIS additive model {\tt laor} \citep{1991ApJ...376...90L}. We
allowed the emissivity index of the model to vary between $2-4$, so
that we got dominant flux from the inner regions of the accretion disk
compared to the outer regions, thus giving rise to a broad
component. We found that the addition of this component did not
improve the fit statistically. {Previous studies on this source by
\cite{2009A&A...507..159D,2004A&A...413..913P,2003A&A...401..903G}
also did not find any broad underlying feature. \cite{2003A&A...401..903G} however detected an emission line at $5.7 \kev$ which he interpreted as the red horn of the FeK line profile. \cite{2010ApJS..187..581S} had studied the \chandra{} data of the source and found a narrow FeK$\alpha$ line at $6.4\kev$.} We also found weak
positive residuals at $7.05 \kev$ which when modeled with a narrow
Gaussian improved the fit by $\dc=-17$ for 3 new parameters to
$\cd=196/190$. The narrow absorption line when modeled
with an inverted narrow Gaussian, improved slightly the fit to
$\cd=189/187$ with a line centre energy of $ 7.45_{-0.07}^{+0.07}$. The fit seemed satisfactory in the $2-10 \kev$ regime
which we extrapolated to the lower energies.

On extrapolating the above best-fit model to the whole band ($0.3-10
\kev$), we found a clear presence of soft X-ray excess emission below
$2\kev$. The origin of such soft excess in type 1 AGNs is still
unclear. Several models such as single or multiple blackbodies,
multicolor disk blackbody, blurred reflection from partially ionised
material, smeared absorption, and thermal Comptonization in an
optically thick medium can provide statistically good fit to the
observed soft excess
\citep{1998MNRAS.301..179M,2002MNRAS.331L..57F,2004MNRAS.349L...7G,2004A&A...422...85P,2006MNRAS.365.1067C,2007ApJ...671.1284D}. We
fitted the soft excess with a simple phenomenological {\tt bbody}
model and still found some positive residuals in the soft X-rays for which we used one more {\tt bbody} model and the fit statistics improved from $\rm \chi^2/dof=
1806/257$ to $\rm \chi^2/dof= 382/253$. The model used in ISIS terminology
is {\tt
  wabs$\times$(bbody(1)+bbody(2)+powerlaw+Gaussian(1)+ \\ Gaussian(2)-Gaussian(3))}. {The
best fit blackbody temperatures obtained were $\rm kT_{BB} =0.08 \pm
0.02 \kev$ and $\rm kT_{BB} =0.17 \pm 0.07 \kev$. The fit statistic of $\chi^2_{\nu} \sim 1.51$ was unacceptably high which may possibly be due to a cold reflection continuum present at energies $>4 \kev$. This continuum emission originates from blurred reflection of hard X-ray photons off a neutral medium distantly located from the blackhole, possibly in the torus or the broad line
region. We used the
PEXRAV model \citep{1995MNRAS.273..837M} which gives us the direct
powerlaw emission as well as the reflection from the cold disk to model this feature, and that improved the fit by $\dc=-118$ to $\cd= 264/250$.  The best fit parameters are listed in
Table~\ref{basic-model}. Given the small band pass of the \xmm{} in the
hard X-ray range, the relative reflection
coefficient (R) of {\tt PEXRAV} however could not be properly
estimated, with a value of $2.27_{-0.5}^{+0.7}$. }

The neutral and narrow iron K$\alpha$ line and Compton reflection
from neutral material could physically arise from the same
component. Therefore a single model describing them together would
better constrain the parameter values. We therefore removed the cold
reflection model {\tt PEXRAV} and the Gaussian line at $6.4\kev$, and
used the model {\tt PEXMON} \citep{2007MNRAS.382..194N} that combines
the cold reflection {\tt PEXRAV} with self-consistently generated Fe
K$\alpha$, Fe K$\beta$, Ni K$\alpha$, Fe K$\alpha$ and the Compton
shoulder. To achieve this, the $0.3-10 \kev$ spectrum was fitted with the model
{\tt wabs$\times$(bbody(1)+bbody(2)+powerlaw+PEXMON-Gaussian)}, in ISIS
notation. We ensured that {\tt PEXMON} models only the
reflection component and not the direct powerlaw. The $\Gamma$ and the
norm of {\tt PEXMON} were tied to the corresponding parameters of {\tt
  powerlaw}. The best fit value of the reflection coefficient R is
$-1.16_{-0.20}^{+0.31}$ and the Fe abundance is $0.27_{-0.07}^{+0.10}$ which
seems to be well constrained.

At $7.40 \kev$ we have earlier detected an absorption line. Such a
line had been identifed as highly ionised absorption from Fe by
several authors e.g, \cite{2003ApJ...595...85C,2003MNRAS.346.1025P,2007ApJ...670..978B}. We removed the inverted Gaussian and used a warm
absorber table model generated using a photoionisation code {\tt
  XSTAR} \citep{2004ApJS..155..675K}, which improved the fit by
$\dc=-8$ for 3 extra parameters to $\cd=273/253$. The ionisation parameter derived from the best fit was $\log\xi=
2.87_{-0.14}^{+0.05}$, the column density was $5.15_{-4.0}^{+5.1}\times
10^{21} \cmsqi$, and the outflow velocity calculated with respect to
the systemic velocity was $\sim \rm0.1 c$, which shows that we
detect a highly ionised, high velocity outflow. Since the detection is weak, we carried out a Monte-Carlo simulation to test the significance of this absorption feature. The Monte-Carlo test suggests that the absorption features are detected at $<2\sigma$ significance. \cite{2010A&A...521A..57T} did not detect the absorption features in the same dataset as they have adopted a criteria where only those features with Monte Carlo derived confidence levels $\ge 95\%$ are selected.

We also tested for the
presence of warm absorbers in the energy band $0.3-2 \kev$ using a low
ionisation {\tt XSTAR} warm absorber model. This model was created
assuming a powerlaw ionising continuum and a turbulent velocity of
$200 \kms$ for the cloud, and spans the parameter space: $ -4 \le
\log\xi \le 4$ and $ 19 \le \log\nh \le 23$. We did not detect any
statistically significant warm absorption in the soft X-rays. However
we find a weak emission line which when modeled using Gaussian profile
improves the statistics by $\dc=-17$ to $\cd=256/250$. The line is
narrow with a central energy of $0.532 \pm 0.007 \kev$ in the rest
frame. We analyse in detail the nature of this emission line using RGS
data below. The lack of warm absorption in the soft X-rays makes
ESO198--G024 a good candidate to study the soft-excess emission
spectroscopically in detail as it is relatively unmodified.

\subsection{Describing the soft-excess with physical models}
\label{sub-physical-models}

The two main physical models in vogue which describe the soft-excess
are the blurred reflection from a partially ionized accretion
disk \citep[{\tt reflionx}; ][]{2005MNRAS.358..211R,2006MNRAS.365.1067C} and
the intrinsic Comptonized emission from an accretion disk \citep[{\tt optxagnf};][]{2012MNRAS.420.1848D}. We have used these
models separately in two instances to describe the soft-excess
component.

\begin{figure*}
  \centering \vbox{
    \includegraphics[width=6cm,angle=-90]{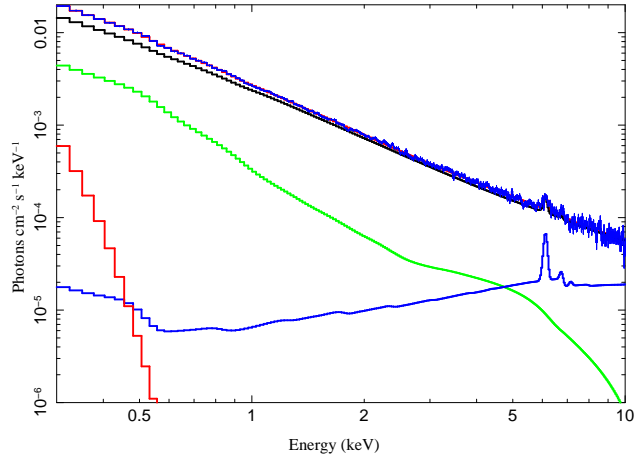}
  }
  \caption{Shows the data and the contributions of the different model
    components when {\it reflionx} was used to model the
    soft-excess. The data and the best fit model are at the top. The
    black line just below it shows the powerlaw absorbed by Galactic
    neutral hydrogen. The green curve shows the contribution of
    the blurred reflection from an ionised disk (reflionx). The blue
    curve with narrow emission lines shows the contribution from the
    PEXMON which is the reflection from a neutral disk. The red curve
    in the extreme left is the contribution from the disk-blackbody.
  }
  \label{model-real}
\end{figure*}

The model {\tt reflionx}
\citep{2005MNRAS.358..211R,2006MNRAS.365.1067C} describes the
soft-excess as a Compton reflection from an ionised disk. It assumes a
semi-infinite slab of optically thick cold gas of constant density,
illuminated by a powerlaw producing a reflection component including
the fluoresence lines from the ionised species in the gas. We blurred
it using {\tt kdblur} to account for the Doppler and gravitational
effects. The full band ($0.3-10 \kev$) was thus fitted with {\tt
  wabs$\times$xstar$\times$(Gaussian+powerlaw+pexmon+\\kdblur(reflionx))}. The
best fit ionisation parameter is $\rm \xi
=493_{-217}^{+116} \xiunit$ and the iron abundance with respect to solar is
$0.77_{-0.21}^{+0.30}$. The fit statistic was $\cd= 271/252$. The narrow emission line modeled using a Gaussian was not
required statistically and hence we removed it. We however detected a
clear excess in the region $<0.5 \kev$. The excess could be 
thermal emission from the thin accretion disk as predicted by
\cite{1973A&A....24..337S}. We thefore used a disk-blackbody component to
model the excess emission, to find that the fit impoved by $\dc=-8$ to $\cd= 263/250$. The F-test significance on addition of
this component is $>90 \%$ and the fit is acceptable. The best fit
temperature of the blackbody is $26_{-4}^{+3} \ev$. This temperature
nearly conforms with that calculated for the innermost radius of thin
disk accreting around a blackhole of mass $\rm M_{BH}\sim 10^8 \times M_{\odot}$ \citep{1999MNRAS.307...41R}. The best fit parameter values
obtained for this set of models are reported in Table~\ref{real-model}, model 1 (column 3). { Figure \ref{model-real} shows the contribution of the various model components in the broad band spectrum.}

The {\tt optxagnf} model, proposed by \cite{2012MNRAS.420.1848D},
describes the soft-excess in terms of color-temperature corrected disk
emission and Compton upscattering of this disk emission by the low
temperature, optically thick inner regions of the disk itself. This
model also describes the hard power-law component by thermal
Comptonization of disk emission by an optically thin, hot corona
external to the disk.
In the {\tt optxagnf} model, the three components -- the thermal
emission of the disk, cool Comptonization in the optically thick disk,
and the hot Comptonization in an optically thin corona, are combined
together assuming that they are all ultimately powered by
gravitational energy released in the accretion process.  We therefore
excluded the blackbody and the powerlaw components from our earlier
best-fit model when using this model.  The parameter $\rm f_{pl}$
determines the fraction of the seed photons that go into producing the
hard X-ray powerlaw. We have frozen the norm of {\tt optxagnf} to one,
since the flux is completely calculated by the four parameters: the
blackhole mass $\rm M_{BH}$, the spin of the blackhole, the relative
accretion rate $\rm L/L_{Edd}$, and the luminosity distance of the
source $\rm D_L$. { We have fixed the value of $\rm log(M_{BH}/M_{\odot})
= 8.1$ \citep{1999MNRAS.307...41R} and $\rm D_L$ to $\rm 192\, Mpc$ (obtained from NED)}.  The spin parameter was set to vary. We obtained the best fit relative accretion rate, $\rm
log(L/L_{Edd})=-1.361_{-0.004}^{+0.003}$. The spin could not be constrained using this dataset. The full band best fit model
stands as {\tt $\rm wabs\times xstar\times (optxagnf+pexmon)$}. As we
had noted that there is no separate powerlaw component, as in previous
cases, to which we could tie the {\tt PEXMON} norm and $\Gamma$, we
had followed the following steps. We had fixed the powerlaw contribution of {\tt optxagnf} model to zero by setting the parameter $\rm f_{PL}=0$ and added a powerlaw model with the model mentioned above. The best fit parameter values of the powerlaw norm and $\Gamma$ thus obtained were noted. We removed the powerlaw from the model and allowed the parameter $\rm f_{PL}$ of {\tt optxagnf} model free to vary. We fixed the pexmon norm and $\Gamma$ to those obtained using the powerlaw model and carried out our fit.

\subsection{The RGS spectral analysis}
\begin{figure}
  \centering

  \includegraphics[width=7cm,angle=-90]{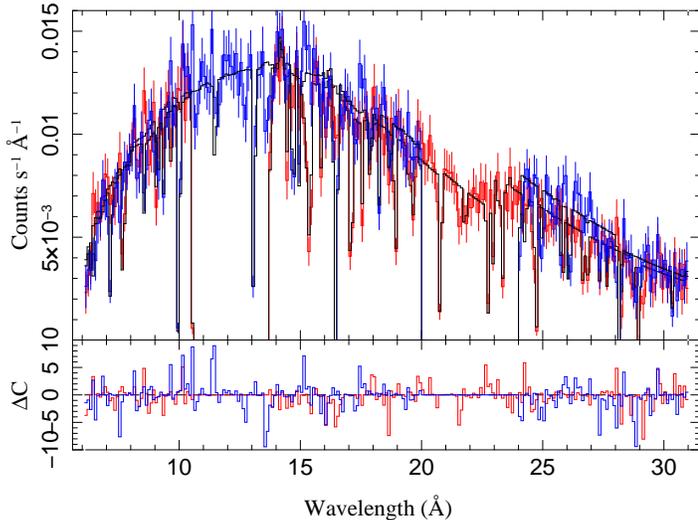}
  \caption{ The observed RGS1 and RGS2 spectral data along with the
    best-fit model (in black) {for observation 3}. There are no statistically significant
    warm absorbers. However there are two narrow emission lines of
    neutral Oxygen and Ne arising from a medium with an ouflow
    velocity of $\sim 12,000 \kms$ at wavelengths $\rm 14.5 \AA$ and
    $\rm 23.4 \AA$. }.
  \label{rgs}
\end{figure}

We fitted the RGS1 and RGS2 spectra simultaneously with the EPIC-pn
spectrum for the purpose of testing the presence of warm absorption
and emission in the high resolution spectra. We used one of the
previously obtained best fit physical models {\tt
  wabs$\times$xstar$\times$(optxagnf+pexmon)} for our continuum. {The
RGS spectra were not grouped and therefore we used the C statistics
\citep{1979ApJ...228..939C}. The best fit value of the statistic for
the simultaneous fit was C/dof$=5250/5215$. We checked for the presence of any ionised absorption feature in the RGS data using the {\tt XSTAR} warm absorber model, but could not find any within $90\%$ statistical significance. This corroborates with our non-detection
of a low ionisation warm absorber component in the EPIC-pn
spectrum. Earlier studies on the source by
\cite{2004A&A...413..913P} and \cite{2003A&A...401..903G} have also not detected
any low ionisation warm absorbers.} However, we find the presence of an
emission feature at $\sim 0.532 \kev$ (rest-frame) in the EPIC-pn data as well as in the RGS spectra. This emission feature may arise from
an ionised cloud intrinsic to the source. We used a Xstar warm emitter
additive table model to estimate the cloud parameters which give rise
to these features. This table model was created assuming a powerlaw
ionising continuum and a turbulent velocity of $200 \kms$ for the
cloud, and spans the parameter space $ -4 \le \log\xi \le 4$ and $ 19
\le \log\nh \le23$. The fit improved by $ \Delta C=-36$ to a
statistic C/dof$=5214/5211$ on addition of this component. The best
fit parameters are: $\log\xi= -1.73_{-0.3}^{+0.4}$, $\nh^{WE}= (9.54\pm
0.5) \times 10^{19}\cmsqi$, $v_{outflow} \sim 12,000 \pm 500 \kms$. The line
was identifed with the neutral Oxygen $K \alpha$ emission line. We
also identify one more emission line modeled by the same warm emitter
component which corresponds to neutral Neon $K \alpha$ at $\sim 0.85
\kev$. Figure \ref{rgs} shows the RGS1 and RGS2 spectra with the best
fit model {\tt$\rm wabs\times Xstar_{WA} \times
  (Xstar_{WE}+optxagnf+pexmon)$} when simultaneously fitted with the
EPIC-pn data, along with the residuals.

\subsection{The Optical Monitor data}
\label{sub:OM} 
\begin{figure}
  \centering

  \includegraphics[width=8cm,angle=0]{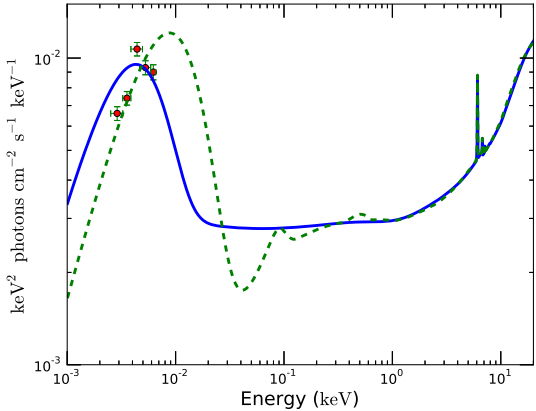}
  \caption{ The two sets of best fit models {\tt reflionx+diskbb} (dashed green) and {\tt optxagnf} (solid blue) obtained from joint fitting of OM and EPIC-pn data. The OM data points are plotted in red. See section \ref{sub:OM} for details.}.
  \label{OM-pn-sed}
\end{figure}

{From Table~\ref{real-model} we find that both the soft-excess models ({\tt reflionx} and {\tt optxagnf}) yield similar fitting statistics to the EPIC-pn data. We further investigated this result by simultaneously fitting the UV data from the optical monitor (OM) telescope and the EPIC-pn data. The OM observed
  ESO~198--G024 simultaneously with the EPIC-pn with five
  filters. We used the SAS task {\tt omichain} to reprocess the OM data. We obtained the flux values from the source list files that were generated by the {\tt omichain} task. The right ascension and declination of ESO~198--G24 were matched in the combined list and we obtained the source flux at the five wavelengths corresponding to the five filters.  The Galactic extinction correction was done following \cite{1999PASP..111...63F} redenning law with $\rm R_v= 3.1$. The observed flux at $2120\rm \AA$ (UVW2 filter) is $5.23\pm0.06\times 10^{-15} \funit \AA^{-1}$ and after correcting for the Galactic reddening ($\rm A_{\lambda}=0.282$) we obtained a flux of $6.78\pm0.07\times 10^{-15} \funit \AA^{-1}$. The Galactic extinction corrected fluxes obtained in the five filters are given in the Table \ref{OM}.

  The measured OM flux at different wavelengths are also affected by the host galaxy contamination, the nuclear emission lines and the intrinsic reddening, which yield a considerable amount of systematic uncertainty in the measured optical-UV continuum flux. Accurate measurements and estimates of these quantities are not known, so we have added a typical $5\%$ systematic error to the OM fluxes. The two sets of models {\tt reflionx} and {\tt optxagnf}, were used to jointly fit the OM and EPIC-pn data. The best fit parameter values are quoted in Table \ref{real-model} (columns 5 and 6). We find that both the models describe the data with similar statistics. The higher values of $\cd$ are due to the small error bars on the OM data points, so even a small deviation from the model yields a poor statistic. Note that the parameters of the models {\tt pexmon}, {\tt Xstar-WA} and {\tt Gaussian} were frozen in the joint fit as they were determined by the EPIC-pn fit only. Figure \ref{OM-pn-sed} shows the plot for the two best fit models along with the OM data points.}  
  
\subsection{The long-term X-ray and UV variability}
\label{sub:variability} 
{The joint fit to the OM and the EPIC-pn data from the observation 3 did not prefer any one of the soft-excess models over the other. {We then looked for possible variability in the source and its pattern to help distinguish between them.} We also performed spectral analysis of observation 1 (2000) and observation 2 (2001) (see Table \ref{obs}). We fitted the three EPIC-pn ($0.3-10\kev$) datasets separately with simple models $\rm wabs*(bbody+powerlaw+pexmon+Gaussian)$. The powerlaw slope and the blackbody temperature for all the three datasets were similar. However the X-ray flux in the $0.3-10 \kev$ band had varied from observation 2 to 3 (see Figure 6 and Table \ref{Xray-flux}). The EPIC-pn datasets for observations 1 and $2$ were analysed in detail using the two models {\tt reflionx} and {\tt optxagnf} for the soft-excess which are reported in Table \ref{dataset-1-2}. For observation 1 both models describe the dataset well, but in case of observation 2 the {\tt reflionx} model gives a statistically better fit ($\cd=229/232$) than the {\tt optxagnf} ($\cd=253/235$). The continuum model parameters except for the normalisations are similar for the two observations (see Table \ref{OM}). In the UV band the flux increased by $23\%$ from the observation 2 to observation 3. The first observation used only one filter (U band) which was not used in the second observation. We also find that when the UV flux increased by $23\%$, the X-ray flux decreased by almost the similar fraction ($\sim 20\%$). { We have also carried out simultaneous fit of the EPIC-pn and OM datasets for the three observations to obtain better constraints on the parameters for the lower signal to noise data. We found that the parameter values for the two sets of models {\tt reflionx} and {\tt optxagnf} are similar to those obtained in individual fits and therefore do not quote them separately.} We discuss the implications of these results in the next section.}



\begin{figure}
  \centering \label{Xray-UV-var}
  \includegraphics[width=6cm,angle=-90]{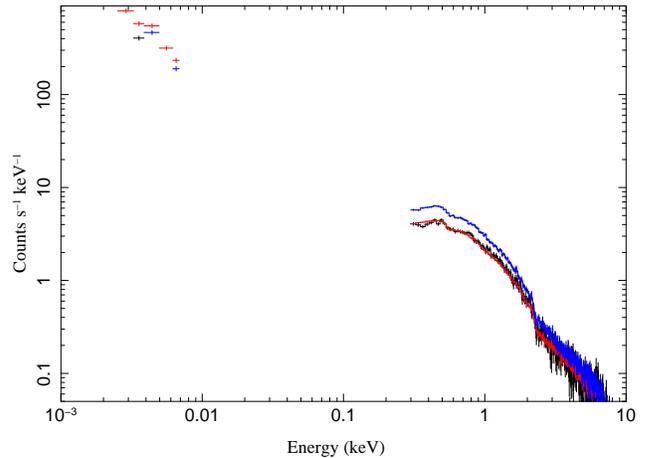}
  \caption{The OM and EPIC-pn data are plotted for the three \xmm{} observations. The colors black, blue and red stands for the observations 1 , 2 and 3. We see flux variability between the observations in both the EPIC-pn and OM datasets.}.
  \end{figure}

\section{Discussion}

The broad band EPIC-pn spectrum of ESO 198--G024 is well described by
a soft-excess component, an iron line complex ($6-8\kev$), and neutral
reflection from distant cold matter. In the iron K band, we detected
neutral K$\alpha$, K$\beta$ emission lines and absorption features at energies
$>7 \kev$. The $0.3-10\kev$ unabsorbed flux of the source is $2 \times
10^{-11} \funit$. {The continuum parameters for e.g. powerlaw slope ($\Gamma =1.81\pm 0.01$), black body temperature ($kT=0.08\pm 0.02, \, 0.17\pm 0.07 \kev$) and the neutral reflection coefficient ($R=-1.62\pm0.1$) are similar to those found in typical Seyfert 1 galaxies \citep{2005A&A...432...15P, 2012ApJ...745..107W}.} Below we discuss our main results.

\subsection{The Soft X-ray excess emission}

ESO~198--G024 shows a prominent soft X-ray excess emission below
$<2\kev$ over an absorbed power-law component. The soft excess flux is
$\sim 1.00 \times 10^{-12} \funit$, which is $\sim 5\%$ of the total $0.3-10 \kev$ unabsorbed flux. Two blackbodies
with temperatures $\rm kT_{BB} = 0.08\pm 0.02 \kev$ and $\rm kT_{BB} = 0.17\pm 0.07 \kev$ yielded satisfactory fit for the
soft excess. Thus, the strength and the temperature of the soft excess
from ESO~198--G024 are similar to those observed from other Seyfert 1
galaxies. We found no significant warm absorption in the $0.3-2 \kev$
band. Since the soft-excess was not modified by complex warm
absorption, we were able to investigate in detail its possible origin.
The temperature of the soft excess when described as thermal emission
is much higher than that predicted from an optically thick and
geometrically thin accretion disk. To spectrocopically ascertain the
possible origin of the soft-excess we modeled it with two physical
models -- the blurred reflection {\it reflionx} and intrinsically
Comptonized disk emission {\it optxagnf}. Both the models describe the soft-excess
statistically well. We discuss below the possible implications of each
soft-excess model for ESO~ 198--G024. The best fit parameter values
are quoted in Table \ref{real-model} columns 3 and 4.

The ionisation parameter of the reflecting disk of the {\it reflionx}
model is $\xi =493_{-217}^{+116} \xiunit$, suggesting a moderate
ionisation state. The best fit inner radius $\rm <2.5 R_G$, and an
emissivity index of $4.23_{-0.02}^{+0.50}$ of {\it kdblur} clearly
indicate that major part of the soft excess flux is emitted from a
region very near to the central accreting blackhole. This gives us a
picture of an accretion disk which is ionised and reflects the
powerlaw continuum incident on it, and most of this reflection comes
from within a few gravitational radii. We required an additional disk black
body component to model the excess below the energy range $0.5 \kev$,
with a best fit temperature of $\rm kT_{BB}= 26_{-4}^{+3} \ev$. This points to a dual origin of the soft excess, i.e, the accretion disk acts as a thermal ionised material which emits a blackbody spectrum, and also emits reprocessed fluorescent spectrum. In Fig. \ref{model-real} we see the individual contribution
of each of the continuum and discrete components in the spectra. The
{\tt reflionx} flux in the $0.3-2 \kev$ band obtained is $2.23\pm 0.06
\times 10^{-12} \funit$ and the disk-blackbody flux in the same energy
range is $7.76\pm 0.06 \times 10^{-14} \funit$. Thus the disk-blackbody
component is weak and contributes just $\sim 3.5 \%$ of the total
soft-excess flux. 

The {\it optxagnf} describes the soft-excess in terms of the intrinsic
thermal emission from the disk and the thermal Comptonization in the
disk itself \citep{2012MNRAS.420.1848D}. This model also includes the
powerlaw component extending to high energy in the whole band of
$0.3-10 \kev$. The best fit value of the parameter $\rm f_{PL}
=0.80\pm 0.02$ points to the fact that $\sim 80 \%$ of the power
released in the accretion process is converted to powerlaw
photons. The best fit value of the temperature of the thermal
electrons which Comptonise the seed photons to produce the soft-excess
is $\rm kT_e = 240_{-20}^{+40} \ev$ with an optical depth of $15.76$.

{ In type 1 AGN, the two models {\tt reflionx} and {\tt optxagnf} have been claimed to explain the soft X-ray excess.} The
blurred reflection model naturally explains the similarlity of the
soft-excess shape observed across various AGN, the shape does not
scale either with the black hole mass or luminosity
\citep{2006MNRAS.365.1067C}. The reflection model also naturally
explains the lag observed between the hard X-ray band dominated by the
direct X-ray continuum and the soft X-ray band dominated by the soft
excess. The observed lag of about $30{\rm~s}$ is interpreted as the
reverberation lag between the direct primary continuum and the
reflection from the inner disk
\citep{2009Natur.459..540F,2011MNRAS.416L..94E}. The reflection model
also explains the remarkable X-ray spectral variability observed in some
AGN where the soft X-ray excess remains nearly constant while the
power-law component varies by large factors, see
e.g. \cite{2005MNRAS.361..795F}.

\cite{2007ApJ...671.1284D} have investigated the time delay between
the emission in the soft and hard X-ray bands {for the sources MRK~1044 and AKN~564}. They found the soft
X-rays leading the hard X-ray band contrary to the expectations from
the reflection model but generally consistent with the optically-thick
thermal Comptonization model. However, it should be noted that in the
reflection model, the soft excess consists of numerous emission lines
blended and blurred due to the high velocities and strong gravity in
the innermost regions. These lines are usually generated by absorption
of continuum photons with energies just above the line energies. The
reverberation delay between the continuum and the blurred reflection
is expected to be short ($\sim 30{\rm~s}$) as observed in some narrow
line Seyfert 1 galaxies
\citep{2009Natur.459..540F,2011MNRAS.416L..94E}. The long time delay
$\sim 1700{\rm~s}$ reported by \cite{2007ApJ...671.1284D} could be
dominated by the delay due to the Comptonization process. In the {\tt
  optxagnf} model of \cite{2012MNRAS.420.1848D}, all the soft X-ray
excess emission is attributed to the disk.  This model is similar to
the thermal Comptonization model {\tt nthcomp} except that it includes the thermal emission from an accretion disk and the inner
region of the disk acts like an optically thick Comptonizing
corona. \cite{2012MNRAS.420.1848D} have shown
that this model describes the soft excess satisfactorily well in the
case of extreme narrow-line Seyfert 1 galaxy RE~J1034+396, as well as
in the low accretion type 1 AGN PG~1048+231. 
\cite{2013MNRAS.431.2441D} have done an extensive study of soft X-ray time lags in AGN. They have found that the time scales of the soft lags are relatively short and are also strongly correlated with the blackhole mass of the AGN, indicating that these lags originate in the inner most regions. Their results best describe the scenario where the delayed soft-excess emission originates from the inner regions of the AGN which is stimulated by a compact central source of hard X-ray photons. ESO~198--G24 is a source in their sample and it has shown a soft time lag of $<2\sigma$ significance. { Confirmation of such a time lag would suggest a reflection origin for the SE. }

{The simultaneously fitted OM and the EPIC-pn data of observation 3 shows both the models {\tt reflionx} and {\tt optxagnf} described the datasets equally well. The larger value of $\chi^2_{\nu}$ is due to the fact that the OM data points have small errors. From the joint fits using the {\tt reflionx+diskbb} model we find that the diskblackbody has a much colder temperature of $kT=3.6_{-0.1}^{+0.2}\ev$. This is comparable with the inner radius temperature of an accretion disk accreting at a rate of $\rm \log(L_{Bol}/L_{Edd})\sim -1.50$ (obtained using {\tt optxagnf} model) which is $\sim 6 \ev$, if we assume an inner most stable radius of six times the Schwarzschild radius $\rm R_S$. The best-fit accretion rate obtained using the {\tt optxagnf} model is comparable to accretion rate $\rm \log(L_{Bol}/L_{Edd})\sim -1.46$ calculated from the broad band $0.001-100 \kev$ SED. The $0.001-100\kev$ bolometric flux calculated from the SED is $1.06\times10^{-10}\funit$.


{ We found flux variability both in the UV as well as in the X-rays. The UV fluxes increased from observation 2 to observation 3 by a factor of $23\%$ (UVW2), while the $2-10\kev$ flux as well as the soft-excess flux decreased by a factor of $\sim 20\%$. This kind of variability appears to be similar to the X-ray variability observed from  black hole X-ray binaries which show a strengthening  powerlaw with decreasing disk emission when they make spectral transitions from the high/soft or thermally dominated state to the low/hard states \citep[see e.g.,][]{1997ApJ...489..865E,2005A&A...440..207B}. The observed UV and X-ray variability from ESO 198-G24 can possibly be explained in the truncated disk scenario with a spherical corona lying between the truncated radius $\rm R_{in}$ and the inner most stable orbit $\rm R_{ISCO}$. If the inner radius of the truncated disk $R_{in}$ comes closer to the $\rm R_{ISCO}$ , the effective area of the disk will increas leading to an increase in the UV flux while the corona will be compressed and hence the hard X-ray emission will decrease. This is also what we observe when we fit the data for all the three observations using the model {\tt reflionx}. We find that the best fit inner radius decreases from $\rm R_{in}=4.93_{-1.10}^{+1.12} \,R_G$ to $\rm R_{in}<2.5\,R_G$ from observation 2 to 3, and we note that the SE flux has also decreased during that time and the UV flux has increased. However it is not clear physically how the inner radius of the truncated disk can be shortened.}

The {\tt optxagnf} model has three distinct emission components-- the UV bump, the soft-excess, and the hard X-ray powerlaw, assuming that they are all powered by the gravitational energy released in accretion. The model assumes that the emission thermalises to a blackbody only at a large radius, and at smaller radii the gravitational energy released through accretion is split between powering the soft-excess (optically thick Comptonised disk emission) and the powerlaw (optically thin corona). The flux variability of the source ESO~198--G24 can be possibly explained with this model if the coronal radius moves nearer to the blackhole. In such a scenario the UV emission will increase with a decrease in powerlaw as well as the soft-excess. However, it is not clear physically how the coronal radius can be shortened.

\subsection{The Fe line complex and the neutral reflector}

In ESO~198--G024, the Fe K$\alpha$ line is slightly broader than the instrument resolution (FWHM velocity $<14000
\kms$). Also there is a weaker but significant Fe K$\beta$ line
detected at $7.05 \kev$ rest wavelength. However there is no
statistically compelling broad Fe K feature at $\sim 6 \kev$, clearly
indicating that the lines arise from a distant region from the central
blackhole. Also the best fit line energy of the Fe K$\alpha$ is
$6.41_{-0.05}^{+0.04}\kev$ which indicates that the fluorescent line is emitted
from a relatively neutral medium.  The neutral reflection model PEXMON
could describe the Fe K emission lines consistently with the reflected
continuum. The reflection coefficient of $\rm R=-1.62_{-0.12}^{+0.10}$
points to an origin from a reflector with a complicated
geometry. A reflection coefficient of value 1 would imply a reflector
which subtends a solid angle of $2\pi$ at the powerlaw emitter, which
is assumed to be isotropically emitting. So in the case of
ESO~198--G024 either the emitter is not isotropically illuminating the
neutral medium, which may be the torus, or the geometry of the
reflecting medium is more complicated. Such values of reflection
parameters are not uncommon \citep{2007MNRAS.382..194N}, and in such
cases a more complicated geometry is assumed. The best fit Fe
abundance is $0.20\pm0.04$ with respect to solar, for the neutral
reflector.

As we noted earlier, the soft-excess of ESO~198--G024 could also be
described by a relativistically blurred reflection model. This
hints at the possiblity of detecting a broad Fe emission line. {However,
we do not detect one, which possibly means that the line is blurred beyond detection.}

\subsection{The high ionisation absorbers}

A highly ionised high velocity warm absorber has been {tentatively} detected in the
energy range $> 7 \kev$ at $<2\sigma$ significance. The best fit ionisation parameter is $\xi
\sim 1000 \xiunit$ points to the fact that these features are mainly
from {Fe}{XXV} and/or {Fe}{XXVI} K-shell resonance absorption having
large column densities \citep{2006AN....327.1012C}. We have detected an
absorption column of $ 5\times 10^{21} \cmsqi$. These outflows are possibly connected with the accretion disk winds, since detailed studies of accretion disk winds by \cite{2003MNRAS.345..657K} and
\cite{2004ApJ...616..688P} suggested that the inner regions of the
outflowing material can be highly ionised by the intense radiation and
can have large outflow velocities. Since this gas is highly ionised we
do not expect it to show any signature in the soft X-ray where the
signal to noise is better. The signatures of these features are only
found for energies $>7 \kev$ where the SNR is not very good, hence the
detections are usually weak.

\section{Conclusion}

We have performed a detailed analysis of a long {\it XMM-Newton} observation of ESO198--G24. The main results are summarised below: \\

\begin{enumerate}

\item The $0.3-10 \kev$ continuum is well described by a powerlaw of
  slope $\Gamma= 1.82$, a soft excess component, and a neutral
  reflection component including FeK emission lines.

\item {The soft excess is well described statistically by two
  physical models, {\it reflionx} and {\it optxagnf}. It
  may either arise from a combined effect of the reflection from an
  ionised disk which is heavily blurred due to gravitational effects, and a
  thermal emission from a thin accretion disk, described by {\it reflionx}. It may also arise from Compton upscattering from an optically thick thermal plasma of temperature $kT_e= 0.24\kev$, described by {\it optxagnf}. Jointly fitting the OM and EPIC-pn data we found that both the models yield similar fit statistics and hence cannot be favored one over the other. }
  
 \item 
 {Variability found in the observations 1, 2 and 3 of the source show that the soft-excess flux decreases when the UV flux increases by the similar amount. This observation can be described in the {\tt reflionx} scenario by a truncated accretion disk whose inner most radius has come closer to the blackhole leading to an increase in the UV flux and subsequently compressing the corona leading to reduction of the powerlaw flux and therefore soft-excess flux. We find that the best-fit inner radius of the accretion disk decreases from $\rm R_{in}=4.93_{-1.10}^{+1.12}R_G$ to $\rm R_{in}<2.5R_G$ from observation 2 to 3. Possibly the {\tt reflionx+diskbb} model describes the soft-excess better for the source ESO~198--G024.}

\item We detected the presence of an FeK$\alpha$ line
  with $\rm v_{FWHM}< 14000 \kms$. We also detected a narrow
  FeK$\beta$ line. These lines can arise from the torus or inner
  BLR. We do not detect any broad Fe line component.

\item A neutral reflection component was detected, which also
  consistently modeled the FeK$\alpha$ line suggesting a common origin
  for both.

\item A high ionisation and high velocity warm absorber was tentatively 
  detected at $<2\sigma$ significance. We do not find any evidence for a low ionisation warm absorber.

\item We detect a weak warm emission component with two prominent lines
  identified as neutral O K$\alpha$ and neutral Ne K$\alpha$.

\end{enumerate}

{$Acknowledgements:$ This work is based on observations obtained with XMM-Newton, an ESA science
mission with instruments and contributions directly funded by ESA
Member States and NASA. Authors are grateful to the anonymous referee for his/her comments which improved the quality of the manusript. Author SL is grateful to CSIR, Govt of India, for
  supporting this work. }


\begin{table*}
{\footnotesize
\centering
  \caption{The best fit model parameters, using the basic set of models wabs*(bbody+bbody+PEXRAV+egauss+egaus-egaus).  \label{basic-model}}
  \begin{tabular}{l l l l llll} \hline\hline 

Model &  parameters        &  Values   &Comments/ $\dc$ improvement \\  \hline \\     

wabs & $\rm N_H$ (fixed)  & $3 \times 10^{20} \cmsqi$ \\ \\

bbody &  norm         &  $4.79_{-0.18}^{+0.20} \times 10^{-5}$   &  \\ 
      &$\rm kT_e$ &  $0.08 \pm 0.02\kev$ \\ \\
bbody 2 &  norm         &  $1.13_{-0.30}^{+0.20} \times 10^{-5}$   &  $\dc = -1424$\\ 
        &$\rm kT_e$     &  $0.17 \pm 0.07\kev$                   &  (for the two bbody components) \\ \\

PEXRAV &  norm        &  $0.0028\pm 0.00002$           &   $\dc =- 118$          \\ 
       & $\Gamma$    & $1.82\pm 0.02$ \\ 
       & rel reflection  $\rm ^a$ & $ 2.27_{-0.51}^{+0.71}$  &    \\ 
       & Fe abundance &  $0.58_{-0.20}^{+0.40}$  & \\ 
       &inclination &  60  & pegged at this value (cos$(\theta) = 0.5$). \\ \\

Gaussian1 & norm $\rm ^b$ & $9.65_{-2.00}^{+1.50} \times 10^{-6}$ &  $\dc =- 79 $\\ 
          &Line E &  $6.41\pm 0.05$ \\
          & $\sigma$ &  $0.08_{-0.04}^{+0.05}$ \\ \\

Gaussian2  &norm $\rm ^c$ & $2.7_{-1.20}^{+1.10}  \times 10^{-6}$ &  $\dc =- 17$\\ 
           &Line E &  $7.05\pm 0.05$ \\ \\

Gaussian3 &norm $\rm ^d$ & $2.0 \pm 1.30 \times 10^{-6}$ & $\dc = -7$\\ 
           &Line E &  $7.45_{-0.07}^{+0.07}$ \\ \\ \hline

$\rm \chi^2/dof$ &  $264/250\sim 1.04$ & \\ \\

\hline \hline
\end{tabular} \\ 
}
$\rm ^a${Due to small band pass of XMM in hard X-ray, the relative reflection parameter could not be estimated properly.}\\
$\rm ^b${We have used a broad Gaussian to fit the FeK$\alpha$ emission feature.}.\\
$\rm ^c${This is a narrow Gaussian which modeled the FeK$\beta$ emission feature .}\\
$\rm ^d${This Gaussian modeled the FeK absorption feature.}\\

\end{table*}



\begin{table*}
{\tiny
\centering
  \caption{The best fit parameters when the Soft-excess was modeled using physical models for the observation 3, using EPIC-pn dataset as well as EPIC-pn and OM combined datasets. \label{real-model}}
  \begin{tabular}{l l l l llll} \hline\hline

&  &  \multicolumn{2}{c}{EPIC-pn}  &   \multicolumn{2}{c}{ EPIC-pn+OM$\rm ^a$}  \tabularnewline\hline 

Model  & parameters         &  Model 1$\rm ^c$  & Model 2$\rm ^c$  & Model 1 $\rm ^c$   &    Model 2$\rm ^c$     \\   
components    &                   &                   &                  &   &    \\\hline \\

wabs   &  $\nh \, (\cmsqi)$     & $3 \times 10^{20}(f)$&$3 \times 10^{20}(f)$ & $3 \times 10^{20}(f)$  &$3 \times 10^{20}(f)$  \\ 
       & (frozen)                                                                           \\ \\

Warm absorber   &$\nhwa (\cmsqi)$ & $ <10^{22}  $              &$ 4.87_{-5.30}^{+4.02}\times 10^{21}$& $ 10^{21} (f)$& $ 4.87\times 10^{21}(f)$  \\ 
(XSTAR)       &$\rm log(\xi)$     & $2.77_{-0.21}^{+0.20}$      & $2.92_{-0.12}^{+0.10}$             &  $2.77(f)$ &  $2.92(f)$       \\ 
        &redshift $\rm ^b$        & $-0.057_{-0.005}^{+0.01}$ &  $-0.057_{-0.005}^{+0.01}$       & $-0.057(f)$ & $-0.057(f)$    \\ \\

disk-bbody   &norm           &   $1.29^{+50}_{-0.5}  \times 10^7$ & --- & $8.9_{-0.4}^{+1.1} \times 10^9$    \\ 
        &$\rm kT_{BB} \,(\kev)$ & $0.026_{-0.004}^{+0.003}$ & --- &   $0.0036_{-0.0001}^{+0.0002} $      \\ \\

powerlaw&norm                    & $0.00258 \pm 0.0002$      & ---     & $0.00258(f)$         \\ 
        &$\Gamma$                & $1.84_{-0.03}^{+0.02}$           & ---     & $1.84(f)$          \\ \\

PEXMON  & norm                   & $0.00258$              & $0.00271$                    & $0.00258(f)$      &  $0.00271(f)$           \\ 
        &$\Gamma$                & $1.84$                 &   $1.82$                     & $1.84(f)$           &   $1.82(f)$        \\ 
        &rel reflection(R)          & $ -1.62_{-0.12}^{+0.10}$ & $ -1.43_{-0.20}^{+0.15}$      & $ -1.62(f)$   & $ -1.43(f)$   \\ 
        &Fe abundance            &  $0.20 \pm 0.04$  & $0.24 \pm 0.02$                   &$0.20(f)$   & $0.24 (f)$\\  
        &inclination (degrees)   & $<60$                  & $<60$                        &$ 0(f)$            &$0(f)$  \\ \\

Gaussian& norm                &         & $ 4.64_{+2.0}^{-2.8}\times 10^{-5}$& & $ 4.42\times 10^{-5}(f)$ \\
        & Line E (rest)$\kev$  &        & $0.533\pm 0.007$                  & & $ 0.533(f)$ \\
        & $\sigma(\ev)$       &         &  $ 0.001 \kev(f)$                     & & $ 0.001 \kev(f)$ \\ \\

Reflionx &norm         & $4.48_{-0.08}^{+1.50} \times 10^{-8} $& ---& $7.24_{-1.66}^{+1.70} \times 10^{-8} $ \\ 
        &Fe/solar     & $ 0.77_{-0.21}^{+0.30}$       & --- & $0.82_{-0.20}^{+0.30}$    \\ 
        & $\Gamma$    & $1.84 $                       &  --- &$1.84 $   \\ 
        &$\xi$        & $493_{-217}^{+116}$           & ---  & $374_{-95}^{+230}$         \\ \\

Kdblur  &index        &  $ 2.56_{-0.61}^{+2.51}$  & ---&$ 4.23_{-0.21}^{+0.51}$    \\  
        &$\rm R_{in}\, (r_g)$  & ${<2.5}$                & --- &${<2.5}$   \\
        &inclination (degrees)  &  $<27 $         & --- &$<27 $  \\ \\

optxagnf & norm           &   ---&$ 1$   & ---               &   $ 1$       \\
        & $\rm log(L/L_{Edd})$& ---&  $-1.36_{-0.004}^{+0.003}$  & ---  & $-1.51_{-0.004}^{+0.003}$             \\
        & $\rm kT_e$      &  --- &  $0.24^{+0.04}_{-0.02}$      & --- &$0.19^{+0.04}_{-0.02}$           \\
        & $\tau$        &  ---   &  $15.76\pm 2$                    &  --- &  $19.61\pm 3$              \\
        & $\Gamma$       & ---   & $1.82\pm 0.01$               &--- &  $1.83\pm 0.03$                         \\
        & $\rm f_{pl}$    & ---   & $0.80\pm 0.02$              &  ---  &$0.842\pm 0.022$   \\ 
        & $\rm r_{cor}\, (r_g)$   & ---   & $30\pm 12$           & ---  & $72\pm 22$       \\ 
& $\rm Spin$            & ---                            & $<0.99$               & ---               &  $<0.99$    \\ \hline \\ \\

$\rm \chi^2/dof$ &   &  $ 263/250$ & $260/250$  &  $ 306/263$  & $ 316/264$  \\ \\

\hline \hline
\end{tabular} \\ 
}
$\rm ^a${$(f)$ stands for frozen parameters. These parameters were constrained by EPIC-pn fit only.}\\
$\rm ^b${Redshift as noted in the observer frame. Negative implies a blue-shift}\\
$\rm ^c${Model 1= $\rm \tt wabs\times Xstar_{WA}\times (diskbb+powerlaw+pexmon+kdblur(reflionx))$;  Model 2 =$ \rm \tt wabs\times Xstar_{WA} \times (optxagnf+pexmon)$ }

\end{table*}



\begin{table*}
{\footnotesize
\centering
  \caption{The Optical Monitor data for the three observations.  \label{OM}}
  \begin{tabular}{l l l l llll} \hline\hline 

Filters &  Wavelengths &Galactic &  Corrected Flux ($\rm f_{\lambda}$) &   Corrected Flux ($\rm f_{\lambda}$) & Corrected Flux($\rm f_{\lambda}$)  &\\  
        &  (central)   &  extinction   &    $\rm \times 10^{-15}\funit\AA^{-1}$  &$\rm \times 10^{-15}\funit\AA^{-1}$    & $\rm \times 10^{-15}\funit\AA^{-1}$   \\
 used   &    $\rm \AA$  & magnitude        &   (observation-1)           &    (observation-2)     & (observation-3)             & \\ \hline \\ 

UVW2  & 2120 &  0.282 & ---              & $5.51\pm 0.08 $       &$6.78 \pm 0.07$    \\ \\
UVM2 & 2310  &   0.256 & ---             & ---                   & $6.44 \pm 0.043$  \\ \\
UVW1 & 2910  &  0.170  & ---             & $4.78\pm 0.006$       & $ 5.64 \pm 0.004$ \\ \\
U    & 3440  &   0.146 & $2.45\pm 0.004$ & ---                    & $3.60  \pm 0.002$   \\ \\
B    & 4500  &  0.116  & ---             & ---                    &  $2.29 \pm 0.002$     \\ \\

\hline \hline
\end{tabular} \\ 
}
\end{table*}


\begin{table*}
{\footnotesize
\centering
  \caption{The X-ray fluxes of ESO~198--G24 calculated from the EPIC-pn data for the three observations.  \label{Xray-flux}}
  \begin{tabular}{l l l l llll} \hline\hline 

observation & observation & $\rm F_x(0.3-2)\kev$&$\rm F_x(2-10) \kev$&  $\rm F_x(0.3-10) \kev$ &Soft-excess flux &  \\ 
   number   &    id       &      &       &             & (bbody)   \\  
            &             &  $\funit$  & $\funit$  & $\funit$   &  $\funit$ & \\\hline \\

1 &0112910101  &  $9.44_{-0.10}^{+0.11}\times 10^{-12}$ & $ 1.07_{-0.03}^{+0.03}\times 10^{-11}$ & $2.01_{-0.03}^{+0.03}\times 10^{-11} $ & $1.05_{-0.01}^{+0.01}\times 10^{-12}$ \\ \\
2&0067190101   &  $1.23_{-0.005}^{+0.005}\times 10^{-11}$  &  $ 1.29_{-0.01}^{+0.01}\times 10^{-11}$ & $2.51_{-0.07}^{+0.04}\times 10^{-11} $ & $1.23_{-0.005}^{+0.008}\times 10^{-12}  $ \\ \\ 
3 &0305370101  &  $9.18_{-0.02}^{+0.04}\times 10^{-12}$ &  $ 1.086_{-0.01}^{+0.009}\times 10^{-11}$ &  $ 2.00_{-0.02}^{+0.01}\times 10^{-11} $  &$ 1.00_{-0.01}^{+0.008}\times 10^{-12}$  \\  \\
 
\hline \hline
\end{tabular} \\ 
}
\end{table*}

-----------------------------------------------------

\begin{table*}
{\tiny
\centering
  \caption{The best fit parameters when we modeled the EPIC-pn and OM datasets of observations 1 and 2. \label{dataset-1-2}}
  \begin{tabular}{l l l l llll} \hline\hline 
&  &  \multicolumn{2}{c}{Observation 1} &   \multicolumn{2}{c}{ Observation 2} \tabularnewline\hline 
Model  & paramters         &  Model 1$\rm ^c$  & Model 2$\rm ^c$  & Model 1$\rm ^c$ &  Model 2$\rm ^c$  \\   
components   &                   &        &         &    &       \\\hline \\

wabs   &  $\nh \, (\cmsqi)$     & $3 \times 10^{20}$&$3 \times 10^{20}$ & $3 \times 10^{20}$  & $3 \times 10^{20}$  \\ 
       &  (fixed)                                                                           \\ \\

disk-bbody   &norm                  & $>2  \times 10^7$ & --- & $> 3 \times 10^9$ \\ 
            &$\rm kT_{BB} \,(\kev)$     & $0.025_{-0.019}^{+0.001}$ & --- & $0.003_{-0.001}^{+0.001}$        \\ \\

powerlaw&norm                   & $0.00266^{+0.00003}_{-0.0015}$        & ---    & $0.0032 \pm 0.0002$              & ---         \\ 
        &$\Gamma$               & $1.82_{-0.41}^{+0.08}$      & ---    & $1.86 \pm 0.04$                   & ---        \\ \\

PEXMON  & norm                   & $0.00266$                 & $0.0027$                        & $0.0032$              & $0.0032$     \\ 
        &$\Gamma$                & $1.82$                   & $1.82$                          & $1.86$                & $1.85$       \\ 
        &rel reflection          & $ -0.81_{-0.65}^{+0.60}$  & $ -0.64_{-0.2}^{+0.55}$    & $ -0.85_{-0.21}^{+0.37}$  & $ -0.687_{-0.391}^{+0.372}$  \\ 
        &Fe abundance            & $0.16_{-0.37}^{+0.5}$   & $0.88 \pm 0.5$            &$0.38_{-0.22}^{+0.40}$      & $0.60_{-0.18}^{+0.40}$  \\  
        &inclination (degrees)   & $<60$                    & $<60$                      & $<60$                       & $<60$   \\ \\

Reflionx &norm               & $7.6_{-3.5}^{+3}\times 10^{-9}$  &---                     &  $5.22_{-0.88}^{+1.66}\times 10^{-8}$  &--- \\
        &Fe/solar                & $ <20$                         & ---                    &  $ 0.50_{+0.21}^{-0.20}$                 &---  \\ 
        & $\Gamma$               & $1.82 $                        &  ---                   &  $1.86$                                & ---\\ 
        &$\xi$                   & $<10000$         & ---                    &  $971_{-210}^{+400}$                    &--- \\ \\

Kdblur  &index                  &  $4.23_{-2.10}^{+2.20}$                         & ---  & $4.8_{-1.0}^{+3.5}$ & --- \\  
        &$\rm R_{in}\, (r_g)$   & ${<4.5}$                       & ---                     &  $4.93_{-1.10}^{+1.12}$ & ---\\
        &inclination            &  $<60 $degrees                 &                         &$<60 $degrees            & ---\\ \\

optxagnf & norm                  &   ---                          &$ 1$                     & ---                & $ 1$          \\
        & $\rm log(L/L_{EDD})$  & ---                            & $-1.42_{-0.21}^{+0.55}$&      ---          & $-1.44_{-0.05}^{+0.05}$\\
        & $\rm kT_e$            &  ---                           &  $0.21^{+0.04}_{-0.02}$  & ---               & $0.20^{+0.04}_{-0.02}$\\
        & $\tau$                &  ---                           &  $20.63\pm 3$                 &  ---              &  $19.68\pm 2$      \\
        & $\Gamma$              & ---                            & $1.79_{-0.03}^{+0.07}$   &---                &  $1.85_{-0.04}^{+0.07}$ \\
        & $\rm f_{pl}$          & ---                            & $0.89\pm 0.02$           &  ---              &  $0.851\pm 0.101$        \\
        & $\rm r_{cor}\, (r_g)$ & ---                            & $<100$               & ---               &  $65_{-21}^{+55}$    \\
	& $\rm Spin$            & ---                            & $<0.99$               & ---               &  $<0.76$    \\ \hline

$\rm \chi^2/dof$ &   &  $ 163/183$ & $164/183$  &  $243/232 $  &  $253/232 $ \\ \\

\hline \hline
\end{tabular} \\ 
}
$\rm ^c${Model 1= $\rm \tt wabs\times (powerlaw+pexmon+kdblur(reflionx))$;  Model 2 =$ \rm \tt wabs \times (optxagnf+pexmon)$  }

\end{table*}


\bibliographystyle{mn2e} \bibliography{mybib.bib}

\label{lastpage}

\end{document}